%% file: FOCUS-Framework_whitepaper_DRAFT.tex
\documentclass[twocolumn, notitlepage]{scrartcl}  
\makeatletter

\newcommand{\framework}{FOCUS\,}
\newcommand{\frameworkF}{FOCUS-Framework\,}
\usepackage[T1]{fontenc}	 
\usepackage[utf8]{inputenc}  
\usepackage[english]{babel}	
\usepackage{ifpdf} 
\usepackage{calc}  
\usepackage{lipsum}  
\usepackage{blindtext} 
\usepackage{cuted}  
\usepackage{datetime}
\usepackage{lastpage}
\usepackage{graphicx}
\usepackage{afterpage}
\usepackage{todonotes}					  				  
\graphicspath{{Figures/}{./}} 

\usepackage{geometry}
\usepackage[T1]{fontenc}  
\usepackage{authblk}  

\usepackage[sfdefault]{FiraSans} 

\setkomafont{section}{\firasemibold\color{ercred}\fontsize{14pt}{16pt}\selectfont}
\setkomafont{subsection}{\firasemibold\color{ercred}\fontsize{13pt}{16pt}\selectfont}
\setkomafont{subsubsection}{\firasemibold\color{ercred}\fontsize{12pt}{13pt}\selectfont}
\usepackage{caption}
\usepackage[
]{scrlayer-scrpage}
\clearpairofpagestyles
\ihead{}
\ohead*{\includegraphics[height=8mm]{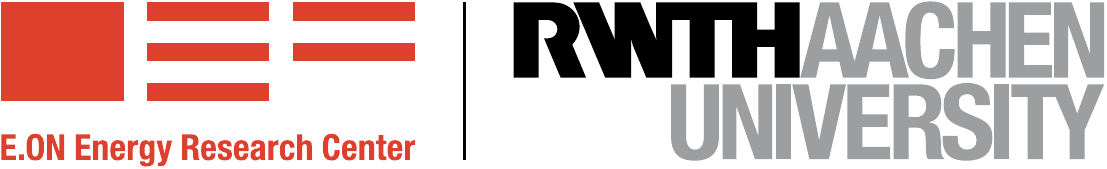}}
\ifoot{}
\ofoot{}
\renewcommand*{\pagemark}{{\usekomafont{pagenumber}{Page \thepage{} of \pageref{LastPage}}}}
\setkomafont{pagefoot}{\footnotesize}
\usepackage{setspace} 
\usepackage{xcolor}
\definecolor{ercred}{RGB}{229,48,39}	
\definecolor{ercblue}{RGB}{16, 88, 176}
\definecolor{ercorange}{RGB}{244, 115, 40}
\definecolor{ercviolett}{RGB}{95, 55, 155}
\definecolor{ercdarkred}{RGB}{155, 35, 30}
\definecolor{ercpink}{RGB}{190, 65, 152}
\definecolor{ercdarkgreen}{RGB}{0, 135, 70}
\definecolor{rwthblue}{RGB}{0,84,159}
\definecolor{rwthblack}{RGB}{0,0,0}
\definecolor{rwthmagenta}{RGB}{227,0,102}
\definecolor{rwthyellow}{RGB}{255,237,0}
\definecolor{rwthpetrol}{RGB}{0,97,101}
\definecolor{rwthturquoise}{RGB}{0,152,161}
\definecolor{rwthgreen}{RGB}{87,171,39}
\definecolor{rwthmaygreen}{RGB}{189,205,0}
\definecolor{rwthorange}{RGB}{246,168,0}
\definecolor{rwthred}{RGB}{204,7,30}
\definecolor{rwthbordeaux}{RGB}{161,16,53}
\definecolor{rwthviolett}{RGB}{97,33,88}
\definecolor{rwthlila}{RGB}{122,11,172}
\colorlet{color0}{ercred}
\colorlet{color1}{ercblue}
\colorlet{color2}{ercorange}
\colorlet{color3}{ercviolett}
\colorlet{color4}{ercdarkred}
\colorlet{color5}{ercpink}
\colorlet{color6}{ercdarkgreen}
\definecolor{color7}{RGB}{0,0,90} 
\definecolor{color8}{RGB}{0,20,20} 

\usepackage{enumitem}  
\setlist{noitemsep} 

\usepackage{listings}

\usepackage{amsmath}
\usepackage{amssymb}
\usepackage{array}

\usepackage{siunitx} 
\usepackage[official]{eurosym}
\DeclareUnicodeCharacter{20AC}{\euro} 
\usepackage{textcomp}					 

\usepackage{float}
\usepackage{placeins}	
\usepackage{caption}	
\usepackage{subcaption}	

\graphicspath{{Figures/}}	
\usepackage[export]{adjustbox}
\definecolor{rotebc}{RGB}{172,43,28}
\definecolor{grauebc}{RGB}{78,79,80}

\usepackage{longtable}	
\usepackage{multirow}	
\usepackage{multicol}	
\usepackage{booktabs}	
\usepackage{colortbl}   
\usepackage{dcolumn} 	
  \newcolumntype{d}[1]{D{.}{.}{#1}}	
  \newcolumntype{C}[1]{>{\centering\let\newline\\\arraybackslash\hspace{0pt}}m{#1}}
\newcolumntype{P}[1]{>{\centering\arraybackslash}p{#1}}

\usepackage[square,sort,comma,numbers]{natbib} 
\usepackage{bibunits}

\usepackage{hyperref} 

\hypersetup{
	hidelinks,
	colorlinks,
	breaklinks=true,
	urlcolor=color2,
	citecolor=color1,
	linkcolor=color1,
	bookmarksopen=false,
	pdftitle={Title},
	pdfauthor={Author},}
	
\usepackage{url}
\usepackage{doi}

\AtBeginDocument{
  \renewcommand{\abstractname}{Abstract}
}

\newlength{\tocsep} 
\usepackage{titletoc}
\contentsmargin{0cm}

\titlecontents{section}[\tocsep]
	{\addvspace{2pt}\small\bfseries\sffamily}
	{\contentslabel[\thecontentslabel]{\tocsep}}
	{}
	{\hfill\thecontentspage}
	[]

\titlecontents{subsection}[\tocsep]
	{\addvspace{2pt}\sffamily}
	{\contentslabel[\thecontentslabel]{\tocsep}}
	{}
	{\ \titlerule*[.5pc]{.}\ \thecontentspage}
	[]

\titlecontents*{subsubsection}[\tocsep]
	{\footnotesize\sffamily}
	{}
	{}
	{}
	[\ \textbullet\ ]

\newcommand{\Abstract}[1]{\def\@Abstract{#1}}
\newcommand{\Keywords}[1]{\def\@Keywords{#1}}
\newcommand{\keywordname}{Keywords} 

\title{\framework: A framework for energy system optimization from prosumer to district and city scale}
\author[1,4]{Jingyu Gong}
\author[2]{Yi Nie}
\author[1,4]{Jonas van Ouwerkerk}
\author[3,4]{Felix Wege}
\author[1,4]{Mauricio Celi Cortés}
\author[3,4]{Christoph von Oy}
\author[1,4]{Jonas Brucksch}
\author[1,4]{Christian Bußar}
\author[2]{Thomas Schreiber}
\author[1,4]{Dirk Uwe Sauer}
\author[2]{Dirk Müller}
\author[3,4]{Antonello Monti}
\affil[1]{Institute for Power Generation and Storage Systems, E.ON Energy Research Center, RWTH Aachen}
\affil[2]{Institute for Energy Efficient Buildings and Indoor Climate, E.ON Energy Research Center, RWTH Aachen}
\affil[3]{Institute for Automation of Complex Power Systems, E.ON Energy Research Center, RWTH Aachen}
\affil[4]{Jülich Aachen Research Alliance, JARA-Energy}
\date{}  
\Abstract{Decarbonizing the energy sector is one of the main challenges to combat the climate crisis. Cities play an important role to reach climate neutrality as more than 70\% of global CO\textsubscript{2} emissions originate from urban areas. Decarbonization of energy supply systems can be achieved through various means, including the use of renewable energy sources, improving the efficiency of technologies, the coupling of different energy sectors, and the use of flexibility considering individual prosumer behaviour. This leads to an increasingly decentralized energy system, which is challenging to operate in a robust and cost-effective way. The evaluation of technologies and subsystems can only be done from the perspective of the system in which it is embedded and it is highly dependent on their networking and application scenarios. Therefore, the design and operation of energy systems require adequate computation and evaluation tools, which offer a holistic view of all interconnected components. The currently available optimization tools have limitations, such as limited scope of technologies and sectors, high requirements on data, high computational cost, and difficulty in handling multi-objective optimization. To overcome these limitations a software framework called \framework for the flexible and dynamic modeling of any urban sector-coupled energy system is developed. The framework includes a library containing models for different technologies and offers a variety of parameter sets for each technology. \framework can handle multi-objective problems by returning Pareto-optimal fronts, which helps users to discover the trade-off between criteria and objectives. The developed tool can identify new flexibility potentials in the energy system, actively support companies in the respective field to optimize urban energy system planning solutions, and determine possible threads to the stable operation of such systems.}

\begin{document}

	\twocolumn[{
		\vskip20pt
		{\raggedright\color{ercred}\sffamily\bfseries\fontsize{20}{25}\selectfont \@title  \par}
		\vskip10pt
		{\raggedright\sffamily\fontsize{12}{16}\selectfont \@author \par}
		\vskip18pt

		\begin{scriptsize}  
			\fcolorbox{ercred}{white}{
				\parbox{\textwidth-2\fboxsep-2\fboxrule}{\centering
					\colorbox{ercred!10}{%
						\parbox{\textwidth-4\fboxsep-2\fboxrule}{
							\ifx\@Keywords\@empty%
								\sffamily{\firasemibold \abstractname}\\ \@Abstract
							\else%
								\sffamily{\firasemibold \abstractname}\\ \@Abstract \\[4pt]
								{\firasemibold \keywordname}\\ \@Keywords
							\fi
						}
					}
				}
			}
			\vskip25pt
		\end{scriptsize}
	}]
	

\makeatother
\tableofcontents 

\input{inhalte/sec01.tex}
\input{inhalte/sec02.tex}
\input{inhalte/sec03.tex}

\phantomsection
\section*{How to Get}
\addcontentsline{toc}{section}{How to Get} 
\label{section:howtoget}

\frameworkF is an open-source tool using the \href{https://opensource.org/licenses/MIT}{MIT license} and can be accessed in \href{https://git-ce.rwth-aachen.de/focus}{Gitlab}.

\phantomsection
\section*{Authorship Contribution Statement}
\addcontentsline{toc}{section}{Authorship Contribution Statement} 

\textbf{Jingyu Gong:} Conceptualization, Methodology, Validation, Software, Resources, Formal analysis, Writing - Original manuscript preparation. \textbf{Yi Nie:} Conceptualization, Methodology, Validation, Software, Resources, Formal analysis, Writing. \textbf{Jonas van Ouwerkerk:} Conceptualization, Methodology, Validation, Software, Resources, Formal analysis. \textbf{Felix Wege:} Methodology, Validation, Formal analysis. \textbf{Mauricio Celi Cortés:}  Methodology, Validation, Software, Resources, Formal analysis. \textbf{Christoph von Oy:} Methodology, Validation, Software, Formal analysis. \textbf{Jonas Brucksch:} Methodology, Validation, Software, Formal analysis. \textbf{Christian Bußar:} Supervision, Funding acquisition, Conceptualization, Writing - Review \& Editing, Project administration. \textbf{Thomas Schreiber:} Conceptualization, Review, Project administration. \textbf{Dirk Uwe Sauer:} Supervision, Funding acquisition. \textbf{Dirk Müller:} Supervision, Funding acquisition. \textbf{Antonello Monti:} Supervision, Funding acquisition.

\phantomsection
\section*{Acknowledgements}
\addcontentsline{toc}{section}{Acknowledgements} 

This work was supported by the funding initiative of "FEN Research Campus - Public-Private Partnership for Innovation" from the Federal Ministry of Education and Research (BMBF). The framework is being developed within its project "InEEd-DC", funding no. 03SF0597.

\phantomsection
\bibliographystyle{unsrt} 
\bibliography{sample.bib}
	
\end{document}

%% file: inhalte/sec01.tex
\section{Introduction}

As global greenhouse gas emissions are still growing and the environmental impact becomes visible, many countries worldwide are accelerating the transition of their energy systems to clean energy. For example, the European Commission issued many regulations in 2021 that together form the 'Fit for 55' package, which commits to climate neutrality by 2050 and sets a target of at least a 55\% net reduction in emissions by 2030 (compared to 1990 levels) \cite{WEO:2021}. More than half of the world's population currently lives in cities. In Europe, this value is even higher, reaching 70\% and is estimated to be over 80\% in 2050 \cite{UNDYB:2015}. Cities also account for over 70\% of global CO\textsubscript{2} emissions according to \cite{WORLDBANKBLOGS:2022}, primarily from use of fossil energies in the transportation and buildings sector. Therefore, the transformation towards sustainable urban energy systems plays a major role for the energy transition. Meanwhile, increasing integration of renewable energy sources can be observed globally \cite{WEO:2021}. For reasons of physical stability of the grids, electricity generation has to match consumption at all times. This requires sources of flexibility like storage technologies to match fluctuations in demand and volatile generation. Furthermore, the rather unstable global situation makes it necessary to highlight energy security \cite{Allam:2022}. In this context, the supply of conventional energy carriers is also characterized by high uncertainty. In addition, new technologies such as direct current (DC) technology and hydrogen gas technology are likely to be promoted in the energy system. Therefore, it can be seen that the energy system will change significantly, and it is increasingly difficult to provide design solutions for such complex energy systems.

For designing this new energy system, dynamic energy system optimization models are becoming more and more relevant since they make the consideration of time-dependent operation characteristics possible and ensure dynamic energy equilibrium even in high volatility situations.

An energy system optimization model is a mathematical model that creates an energy balance and conversion relationship for each time step to find the optimal energy supply solution according to a set target. The optimization framework \framework uses mixed-integer linear programming (MILP). In contrast to other existing tools, which mainly focus on one abstraction level and have a pre-defined objective - usually minimizing cost \cite{ReviewOpenSourceSoftware:2022}\cite{CompareOpenSourceModels:2022}\cite{ReviewModelingTools:2018}, the newly developed software framework covers several levels of the energy system in the end-user area in more detail and allows users to build their own objectives. It adopts a bottom-up approach and provides optimization for energy systems on three levels covering the end supply area. This area mainly consists of consumers and prosumers that can not only consume but also generate usable energy mostly from renewable sources \cite{BMWiProsumer:2016}. Figure \ref{fig:framework} shows the considered three levels from single prosumers over districts towards interconnected city districts. Based on the bottom-up approach, the lowest level, i.e. prosumer level, will be optimized according to a set target first. The subsequent optimization of a single district is based on this prosumer level, using the optimized prosumer results as its building elements and exploiting the remaining flexibility in each prosumer. On the final third level, the districts are interconnected to form a multi-sector urban energy system.

The developed software framework is a power calculation and evaluation tool for the holistic consideration of all interconnected sectors and networked components. Therefore, it is capable of identifying new potentials for energy technologies and hence can support companies active in this domain to make better decisions. Furthermore, it can help identifying boundary conditions in the area of urban planning. Against the background of multi-sector energy systems, an evaluation of novel technologies and business models is meaningful and plausible only in the context of a holistic consideration of all coupled energy sectors.

%% file: inhalte/sec02.tex
\section{A Three-Level Optimization Model}

\subsection{Features}

\begin{figure*}[htbp]
\centering
\includegraphics[width=\textwidth]{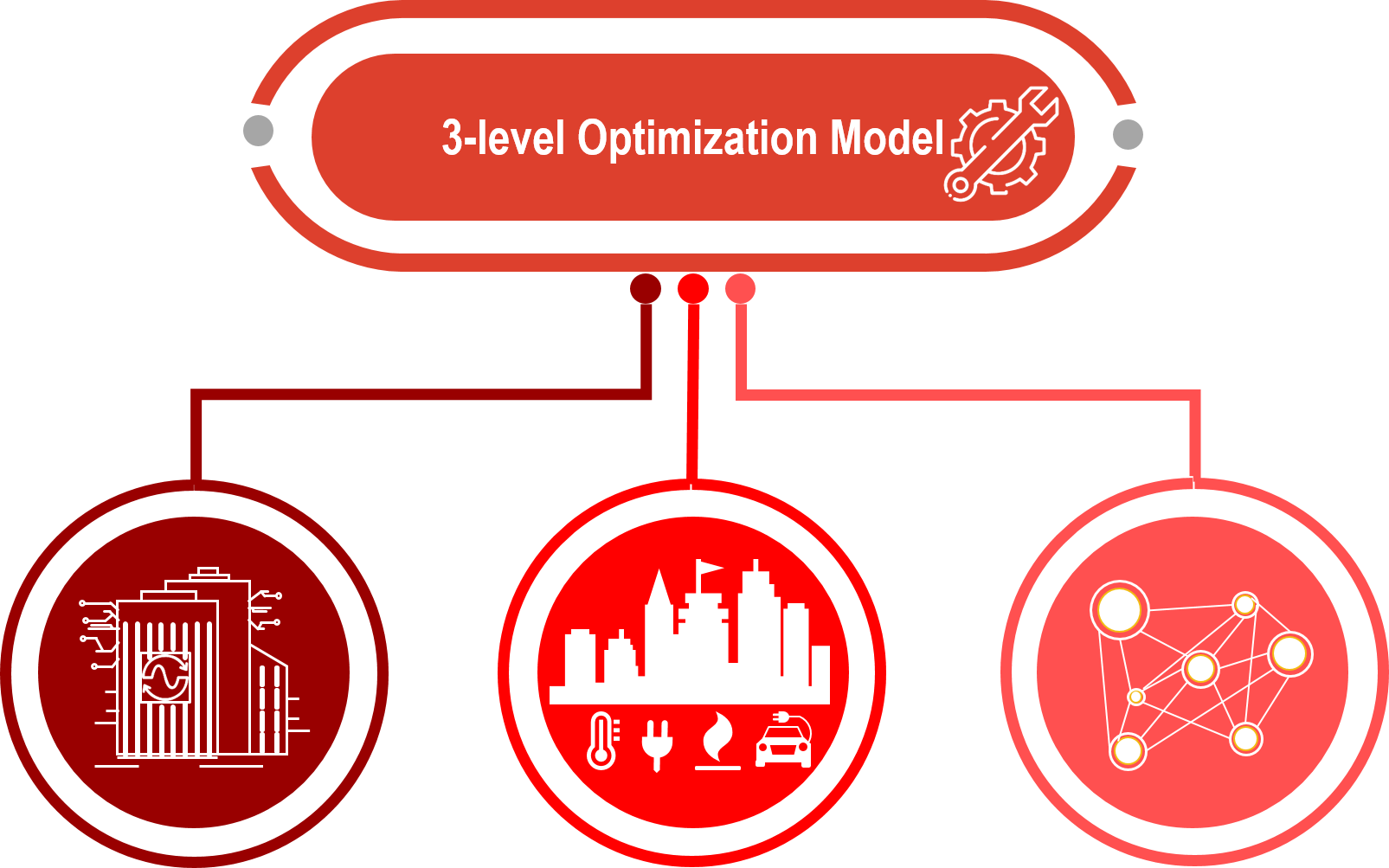}
\caption{\frameworkF}
\label{fig:logoframework}
\end{figure*}

The software framework \framework is programmed in \textit{Python}. Depending on the scope of the optimization, \framework is able to build MILP problems for individual prosumers, districts, and coupled districts. The program adopts a bottom-up approach throughout the entire optimization process. The optimized subsystems from the deeper level are used as building blocks for the systems of the higher level. Through this modular structure, \frameworkF splits complex optimization problems that contain a large number of variables and constraints into multiple sub-problems. It improves computational efficiency, shortens calculation time and also reduces the occurrence of unsolvable problems. Due to this feature, the framework can be easily extended to meet the requirements of different use cases.

Two different kinds of optimization can be generated depending on the received inputs. The first kind is the size optimization for components. In this case, the selection and sizing of energy components are defined as endogenous optimization model variables. As a result, the most suitable sizes for all size-unknown components will be determined through the optimization. The program may even return a size of zero, which indicates a certain technology should not be implemented. The second kind is the operation optimization, in which the component sizing are exogenous model inputs. In this case, the optimization results show the optimal operation in each subsystem for each time step. The two kinds of optimization can be integrated in one problem and solved together by the program.

The framework has the ability to generate multi-objective optimization problems, i.e. problems with more than one optimization objective, for example, to obtain a trade-off between ecological and economic objectives. The optimization results can then be used in a corresponding Pareto front visualization and provide decision-makers with a figure to evaluate different criteria before deciding on a specific investment plan or designing an operational solution.

\subsection{Framework}

\begin{figure*}[htbp]
\centering
\includegraphics[width=\textwidth]{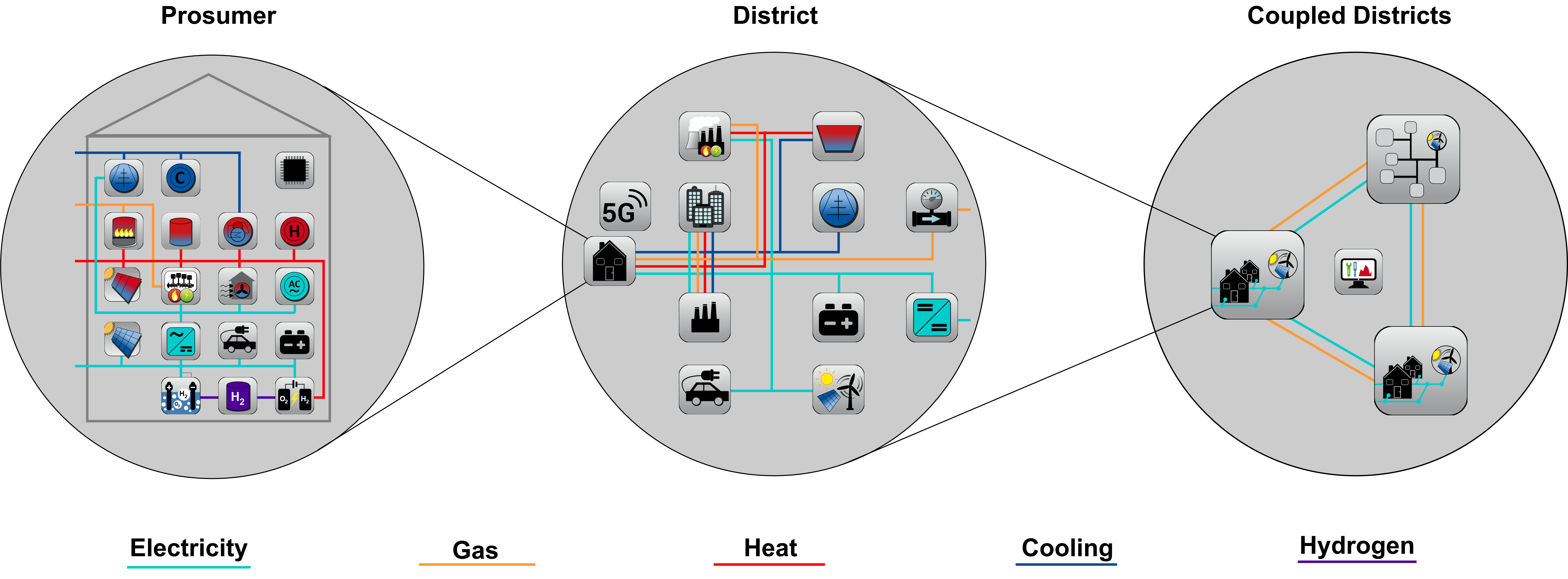}
\caption{Overview of the three-level optimization in the \frameworkF}
\label{fig:framework}
\end{figure*}

The three levels of the model can be abstracted as shown in figure \ref{fig:framework}. On the left side of the figure the basic energy system of a prosumer is highlighted. The prosumer is the smallest unit that can be analysed with a mathematical optimization independently. The optimized prosumers are considered as building blocks for the second level of the framework, i.e. the district model. Besides residential buildings, as shown in the figure, other types of buildings can be modeled as well, such as office buildings with higher demand and different consumption profiles or industrial buildings that may have waste heat from production, among others. In addition to different types of prosumers, traditional energy producers, consumers and central supply infrastructures can also be optimized and added to district models. In the last stage, i.e. the coupled districts, multiple districts are interconnected to build a more complex and broader city energy supply system.

In figure \ref{fig:framework}, the energy components are connected by different colored lines, which represent different energy transfer lines, such as electrical wires for electricity in light blue and water pipes for heat supply in red. The directions of the power flow between components are defined in input files. The connections are usually uni-directional, and only certain devices allow bi-directional flows, such as battery and heat storage. The energy management system (EMS) controls the entire power flows in the prosumer and manages them according to the optimization objective. Currently, users can choose between various objectives: minimizing the annuity, minimizing the operating costs, minimizing the CO\textsubscript{2} emissions, and maximizing the self-consumption. The framework further allows users to easily construct individual optimization objectives by offering access to all parameters and variables during operation.

In the end-user area, prosumers are usually buildings with their own energy generators, such as photovoltaic (PV) generator, wind turbine, and heat pump, among others. They usually utilize storage systems for electricity or heat to increase the self-consumption of generation. As shown in figure \ref{fig:framework}, a prosumer can have multiple demands in electricity, heat, and cooling. To cover the various demands, the prosumer is equipped with generators, storage and distributors from different sectors. Furthermore, the prosumer is able to convert energy between different carriers by using components like heat pumps, gas boilers, electrolyzers, and cooling machines. In \frameworkF, the local energy system of a prosumer is represented by a host \verb|Prosumer| model and its energy components are modeled using different \verb|Component| models. The framework provides ready-to-use parameter sets for the mathematical models of the available energy components, which can be adjusted to the users' needs. Additionally, each \verb|Prosumer| model has a central \verb|EMS| component that controls the power flows.

The optimization of prosumer models in the first layer of \framework provides the sizing of prosumer components and their operation behaviour. This allows to obtain the residual load of each prosumer for the connected grids. The residual load is then used for the second layer of the framework for interconnecting prosumers to form a \verb|District| model. The optimized \verb|Prosumer| models are stored in their host \verb|District| model. The central supply infrastructures such as central battery or heat storage and PV-parks can also be modeled with a \verb|Prosumer| model and added to the host \verb|District| model. Supply networks, i.e. \verb|Electricity grid|, \verb|Heat grid| and \verb|Gas grid|, connect the prosumers and all existing central infrastructures. Analogous to EMS in prosumers, the \verb|District| model has a \verb|Coordinator| that manages the power flows in this district and communicates with the prosumers. The optimization of districts usually aims at maximizing the economic value, by identifying synergies between prosumers and central supply units and exploiting remaining flexibility within each prosumer. Technologies like micro-grids and bi-directional fifth-generation district heating networks can be considered and their potential for future district energy systems can be evaluated. In the third layer of the framework, \verb|District| models are used as building blocks to create the urban energy system model. Large power plants and electricity grids of higher voltage level are added in this layer. The optimization in this layer mainly investigates synergy effects among subsystems and provides the best dimensioning and operation of supply infrastructures and their generation, storage, and transmission components. With this 3-layer optimization approach the model addresses the need for a comprehensive analysis tool to answer research questions in detail on the pathway towards an effective, satisfactory and affordable energy transition from end-user to city scale.

%% file: inhalte/sec03.tex
\section{Summary and Outlook}

The developed \frameworkF addresses the issue of providing energy and power supply solutions, considering interactions between the prosumer, district, and city layer. The component parameter library of the framework covers relevant technologies across all three layers that are relevant to further investigate cross-sector energy supply. The implementation of components is modular and can be easily extended . Moreover, combining a modular topology configuration and the flexible definition of optimization strategies, offers the possibility of modeling both technical relationships and political regulations.

In parallel to the actual development of the underlying models, complexity reduction techniques like a temporal aggregation of the input time series are studied and integrated at the framework level. This approach ensures feasible execution times for large-scale scenario optimization.

The project InEEd-DC is ongoing and is expected to be completed in 2025. Currently, the models of the prosumer layer are developed and can be used directly. Many different types of buildings have been analyzed through different application scenarios. In the later stages of the project, more models of the district layer will be developed, and more subsequent use cases will be analyzed and used to calibrate the models. In addition to developing models for different energy components, new business models are a focus within the development of the framework.